# *NeuCASL*: From Logic Design to System Simulation of Neuromorphic Engines

Dharanidhar Dang, *Member, IEEE*, Amitash Nanda, *Student Member,* Bill Lin, *Senior Member, IEEE,* and Debashis Sahoo, *Senior Member, IEEE*

*Abstract*—With Moore's law saturating and Dennard scaling hitting its wall, traditional Von Neuman systems cannot offer the GFlops/watt for compute-intensive algorithms such as CNN. Recent trends in unconventional computing approaches give us a hope to design highly energy-efficient computing systems for such algorithms. Neuromorphic computing is a promising such approach with its brain-inspired circuitry, use of emerging technologies, and low power nature. Researchers use a variety of novel technologies such as memristors, silicon photonics, FinFET, and carbon nanotubes to demonstrate a neuromorphic computer. However, a flexible CAD tool to start from neuromorphic logic design and go up to architectural simulation is yet to be demonstrated to support the rise of this promising paradigm. In this project, we aim to build *NeuCASL*, an opensource python-based full system CAD framework for neuromorphic logic design, circuit simulation, and system performance and reliability estimation. This is a first-of-its-kind to the best of our knowledge.

*Keywords*— Neuromorphic computing, CAD, on-chip photonics, memristor.

## I. Introduction

In today's era of big data, the volume of data that computing systems process has been increasing exponentially. Deep neural networks (DNNs) have become the state-of-the-art across a broad range of big data applications such as speech processing, image recognition, financial predictions, etc. However, DNNs are highly compute and memory intensive, requiring enormous computational resources. With Moore's law coming to an end, traditional Von Neuman computing systems such as heterogeneous CPU/GPU platforms cannot address this high computational demand, within reasonable power and processing time limitations. Therefore, several unconventional approaches such as spiking neural network, reservoir computing, and neuromorphic computing have been proposed to design next-generation exascale AI accelerators [1].

Among the above-listed paradigms, neuromorphic computing has risen to prominence as alternatives to standard Von Neumann computing systems for many key reasons. It features low-cost and more tightly coupled processing and memory cores that promise low energy consumption and higher degree of parallelism. It targets application areas in classification, control and security that are difficult for more traditional computing systems. In particular, they are well

D. Dang and D Sahoo are with the Department of Computer Science & Engineering, UC San Diego, CA 92093 Email: {dhdang, dsahoo}@ucsd.edu.

A Nanda and B Lin is with the Department of Electrical and Computer Engineering, UC San Diego, CA 92093. E-mail: ananda@ucsd.edu.

suited for convolutional neural networks which is a widely used DNN algorithm for image classifications. Finally, neuromorphic systems offer the ability to leverage alternative technologies as core components. To this end, there is not a single CAD framework that offers neuromorphic logic design, circuit synthesis, and system simulation. This propels us to build a first-of-its-kind python-based full-system CAD toolchain to design next-generation neuromorphic engines.

In this project, we aim to build NeuCASL, a full system CAD framework that offers a broad range of design capabilities starting from neuromorphic logic design to circuit simulation and all the way up to system performance estimation. We plan to make it opensource with an easy-to-use GUI that would benefit the research community.

## II. Ongoing Research

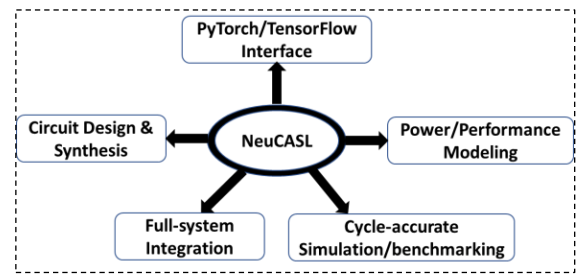

Fig.1: Functional illustration of *NeuCASL*

In its original avatar, "neuromorphic" was referred to custom devices/chips made up of analog components which emulated biological neural activity [2]. Today, neuromorphic computing has broadened to include a wide variety of software and hardware components, as well as materials science, neuroscience, and computational neuroscience research. To support this expansion and make neuromorphic a reality, we plan to build *NeuCASL* in four phases: (i) creating device libraries, (ii) forming power/performance models, (iii) making circuit-level integration toolbox, and (iv) building system design framework. At this point, we are creating python libraries for devices to be used for the neuromorphic systems. The devices include on-chip photonics components such as photonic waveguides, microring modulators (MRMs), semiconductor-optical-amplifiers (SOAs), photodetectors and multi- wavelength laser sources. An MRM is a circular shaped photonic structure with a radius of ~5 µm which is used to modulate electronic signals onto a photonic signal at the transmission source in a waveguide. An SOA is an optoelectronic device that under suitable operating conditions can amplify photonic signals. In a typical high bandwidth photonic link, an off-chip laser source (either on the board or on a 2.5D interposer) generates multiple wavelengths, which are coupled by an optical grating coupler to an on-chip photonic waveguide. Further, we plan to deploy novel memory technologies such as memristor crossbar and photonic phase change memory.

## III. FUTURE WORKS

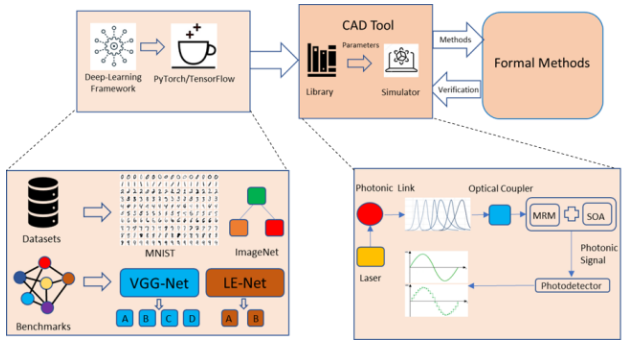

Fig.2: Overview of NeuCASL working principle

We take inspiration from IPKISS [3], a commercial optoelectronic device toolchain to build the circuit simulator for NeuCASL. In the past, we used IPKISS to produce circuit-level parameters for an analog AI architecture design []. However, IPKISS and other state-of-the-art lack system-level integration, fault-check, and formal verification capability. We address them in NeuCASL. Further, we intend to offer circuit synthesis capability in NeuCASL. In the system-integration phase, all of the synthesized components can be combined together to design/implement architecture of choice. We plan to develop a python-based architectural simulator which takes device- and link-level parameters from our circuit-tool, to estimate performance, area, reliability, and energy-efficiency for several benchmarks. For power estimation, we integrate our own power model. At the same time, standard architectural power models such as McPAT [4] and DSENT [5] can be integrated into NeuCASL.

As shown in Fig.2, We plan to use widely adopted deep learning frameworks such as, TensorFlow or PyTorch to train the datasets in conjunction with CAD results of NeuCASL. Each of the benchmarks can be mapped into architectural components programmatically or manually through a GUI interface. This ensures zero pipeline hazards between any two layers of AI or neuromorphic engine developed by *NeuCASL*. The system then determines computational efficiency, energy efficiency, throughput, and prediction error rate to compare the performance with a state-of-the-art AI and neuromorphic accelerators such as PipeLayer [4]. NeuCASL evaluates for the following metrics: *Computational efficiency* represents the total number of fixed-point operations performed per unit area in one second (GOPS/s/mm$^2$); *Energy efficiency* refers to the number of fixed-point operations performed per watt (Giga operations per watt or GOPS/s/W); *Throughput* is the total number of operations per unit time (GOPS/s); and lastly, *Prediction error rate* is the percentage of error in inferring any datasets.

For now, we plan to use two widely used CNN benchmarks: VGG-Net [6] and LeNet [7]. We consider all variants of the VGG and LeNet. For VGG, we use ImageNet dataset [8] having 224×224 images. For LeNet, we use 60,000 224×224 images of MNIST datasets [19] for training and 10,000 224×224 images for testing. Other deep learning benchmarks can also be utilized.

## IV. TIMELINES

| Milestones | 1st Year | 2nd Year |
|---|---|---|
| 1- Performing device-level experiments on standard device tools such as SPICE | ■ | |
| 2- Creating libraries based on device parameters | ■ | |
| 3- Publications of libraries + Open-sourcing | | ■ |
| 4- Building system-level CAD framework for AI architecture | | ■ |
| 5- Publications/Conference presentations of CAD framework | | ■ |
| 6- Online course materials for a global audience | | ■ |

## V. CONCLUSIONS

In this project, we aim to build NeuCASL, an opensource python-based full system CAD framework for neuromorphic logic design, circuit simulation, and system performance and reliability estimation. This is a first-of-its-kind to the best of our knowledge. We plan to build *NeuCASL* in four phases: (i) creating device libraries, (ii) forming power/performance models, (iii) making circuit-level integration toolbox, and (iv) building system design framework. At this point, we are creating python libraries for devices to be used for the neuromorphic systems.

## VI. CONCLUSIONS

This work is supported by National Institute of Health (NIH)/NIGMS grant M138385-01A1 (PI Sahoo).